\documentclass[structabstract]{aa}  
\usepackage{graphicx}
\usepackage{txfonts}
\newcommand{\vsini}{\ensuremath{v_{{\mathrm e}}\sin i}}
\newcommand{\teff}{\ensuremath{T_\mathrm{eff}}}
\newcommand{\logg}{\ensuremath{\mathrm{log}\ g}}

\newcommand{\kms}{\ensuremath{\mathrm{km.s^{-1}}}}
\newcommand{\xit}{\ensuremath{\xi _{t}}}

\begin{document}
   \title{Looking for pulsations in HgMn stars through CoRoT\thanks{The CoRoT 
   space mission was developed and is operated by the French space agency CNES,
   with participation of ESA's RSSD and Science Programmes, Austria, Belgium,
   Brazil, Germany, and Spain.} lightcurves}

%   \subtitle{HgMn}

  \author{G.~Alecian\inst{1}, M.~Gebran\inst{2}, M.~Auvergne\inst{3}, O.~Richard\inst{4}, R.~Samadi\inst{3}, W.~W.~Weiss\inst{5} and A.~Baglin\inst{3} }

  \offprints{G. Alecian}
 
  \institute{LUTH, Observatoire de Paris, CNRS, Universit{\'e} Paris Diderot, 
             5 Place Jules Janssen, 92190 Meudon, France\\
             \email{georges.alecian@obspm.fr}
      \and
	      Departament d'Astronomia i Meteorologia, Universitat de Barcelona,  Mariti i Franques, 1, 08028 Barcelona, Spain \\
%	      \email{mgebran@am.ub.es}
      \and
		 LESIA, Observatoire de Paris, CNRS, 
             5 Place Jules Janssen, 92190 Meudon, France\\
%             \email{michel.auvergne@obspm.fr}
      \and
             GRAAL, Universite Montpellier II, CNRS, Place Eug\`ene Bataillon, 34095 Montpellier, France. \\
%	       \email{richard@graal.univ-montp2.fr}
       \and
             Institut f{\"u}r Astronomie (IfA), Universit{\"a}t Wien,
             T{\"u}rkenschanzstrasse 17, A-1180 Wien, Austria\\
             }

   \date{Received ...; accepted 07 May 2009}

% \abstract{}{}{}{}{} 
% 5 {} token are mandatory
 
  \abstract
  % context heading (optional)
  % {} leave it empty if necessary  
   {HgMn Chemically Peculiar stars are among the quietest stars of the
   main-sequence. However, according to theoretical predictions, these stars
   could have pulsations related to the very strong overabundances of iron peak
   elements, which are produced by atomic diffusion in upper layers. Such
   pulsations have never been detected from ground based observations.
}
  % aims heading (mandatory)
   {Our aim is to search for signatures of pulsations in HgMn stars using the
   high quality lightcurves provided by the CoRoT satellite.
   }
  % methods heading (mandatory)
   {We identified three faint stars ($V>12$), from VLT-GIRAFFE multiobject spectrograph
   survey in a field which was planned for observation by CoRoT. They present
   the typical characteristics of HgMn stars. They were observed by the CoRoT
   satellite during the \emph{long run} (131 days) which started from the 24th
   of October 2007, with the exoplanets CCD's (Additional Programme). In the
   present work, we present the analysis of the ground based spectra of these
   three stars and the analysis of the corresponding CoRoT lightcurves.
   }
  % results heading (mandatory)
   {Two of these three HgMn candidates show low amplitude (less than 1.6  mmag)
   periodic variations (4.3 and 2.53 days respectively, with harmonics) which
   are compatible with periods predicted by theoretical models.}
  % conclusions heading (optional), leave it empty if necessary 
   {}


   \maketitle
%
%________________________________________________________________

\section{Introduction}

HgMn stars are known to form a group among the Chemically Peculiar stars of the
main-sequence (Preston, 1974). Their photosphere is  characterized by strong
abundance anomalies of many metals (see for instance Castelli\&Hubrig 2004).
Overabundances with respect to solar ones can be about two orders of magnitude
and even larger. For instance, Mn can be enhanced by more a factor of 200, and
Hg by a factor up to $10^6$. These stars are found in the range of effective
temperature between about 10000K and  16000K (Pop I, B7-A0 IV-V), they are slow
rotators, often binaries, and no magnetic fields has been yet detected (Shorlin et al. 2002, Wade et
al. 2006). These stars, which are supposed to have no convective motions in
their atmosphere, which do not show any indication of activity, are possibly among the quietest stars of the main
sequence.

The trends of abundance peculiarities observed in HgMn stars are usually well understood in the framework of the atomic diffusion (first proposed by Michaud, 1970). As stated by the classical model, diffusion process builds up abundance
stratifications in the stable atmosphere (and envelope) of HgMn stars, according to the
balance between radiative accelerations and gravity. Theoretical predictions by
Alecian \& Michaud, (1981) for Mn were in excellent agreement with observations
(Smith \& Dworetsky, 1993). However, modelling is not yet able to deal with the large variety of peculiarities observed in these stars. On another hand, detailed studies of stratification build up
remain to be done, because of the heaviness of numerical computations for
optically thin media (see Alecian \& Stift, 2007). Modelling of atomic
diffusion in stellar interior (optically thick case) is more advanced (Richard
et al. 2001). Even if these studies of stellar interiors cannot address the
problem of atmospheric peculiarities, they allow to precise the evolution of
these stars. A recent work  by Turcotte \& Richard (2003) predicts that, due to
iron accumulation in the upper  layers of their envelopes, these stars should at
least be slow pulsators (as the Slow Pulsating B  stars). Presently, there is no
observational evidence of pulsation in HgMn stars, and CoRoT satellite offers an
exceptional opportunity to constrain the models.

A detailed description of the CoRoT mission can be found in the \emph{CoRoT book} (Baglin, 2006). Beside the
two CCD's devoted to asteroseismology\footnote{Each of these CCD's, can measure
photometric variation of 5 bright stars ($m_V<9.5$)}, the exoplanet CCDÕs
photometrically monitor thousands of stars in the magnitude range $10.5\leq
m_R\leq 16$~. This monitoring is designed to detect planet transits but it has
also a high interest for studies of stellar photometric variations. Stellar
science outside the CoRoT core programme have been then allowed to access the
data of exoplanet channel through the \emph{Additional Programme} procedure
(Weiss, 2006).

In Sect. \ref{strategy} we present our strategy to find, through ground based observations (discussed in Sect.~\ref{ground}), faint HgMn stars in fields observed by CoRoT. In Sect. \ref{corotobs}, we present the analysis of the CoRoT data for the targets we identified as HgMn candidates. Finally, we describe theoretical models for HgMn stars, followed by a discussion in view of our observational results.
%__________________________________________________________________

\section{Observational strategy}
\label{strategy}
According to our knowledge of the solar neighborhood, one can estimate that
about 8\% of main-sequence AV-BV stars are HgMn stars. Therefore, applying this
rate to the density of main-sequence AV-BV stars predicted by the
``Besan\c{c}on" model of stellar population synthesis of the Galaxy (Robin et
al. 2003), in the anticenter direction, only about 0.3\% of Pop I stars of any
type may be HgMn. Then, one can expect to have about 35 HgMn stars in a field containing 11400 stars, which is approximately the
number of stars monitored by exoplanet CCD's. The problem is that there is
not any available catalog of such faint HgMn stars, since the MK classification with the dispersion sufficient to recognize the HgMn signature was carried out only for relatively bright stars.

We have then decided to realize prior to the launch of CoRoT\footnote{CoRoT was
launched on 2006 December 27.}, a survey in fields pre-selected for exoplanets
search in the direction of the galactic anticenter, and using VLT-FLAMES/GIRAFFE
(ESO) multiobject spectrograph.

The VLT-GIRAFFE 25' field towards the anticenter direction contains about 490
IV-V stars $11.5\leq m_V \leq 16$, according to the ``Besan\c{c}on" model. The
instrument can provide spectra of 110-120 objects in each field, with good
enough resolution and S/N to identify abundance peculiarities (see next
section). If stars were selected at random, the probability to find HgMn stars
were extremely weak. To overcome this difficulty, we used a preliminary version
of the \emph{Exo-Dat} catalog (Deleuil et al. 2008), where the stellar types in
our fields were roughly determined in view of the CoRoT mission, through
multicolor broad band photometric observations by CoRoT-Exoplanet team. The
\emph{Exo-Dat} catalog consists in 10.6 millions of stars over an area of 220
deg$^2$, observed with the Wide Field Camera (WFC) at the 2.5m Isaac Newton
Telescope (INT) at Roque Muchachos Observatory on La Palma, Canary Islands.
Harris B and V filters and r' and i' ones from the Sloan-Gunn system, were used
(the limiting magnitude is about r' = 20.0). The spectral type, luminosity class
and the mean reddening were obtained by fitting models atmosphere to the
observed fluxes  (Deleuil et al. 2008 and M. Deleuil private communication). All
stars classified as B0-A0 (IV-V), with low contamination\footnote{Only stars
with low contamination are eligible for CoRoT observations. Contamination (by
surrounding sources) is determined in \emph{Exo-Dat} through criteria
established for planets detection.}, were systematically included in our
programme, ensuring about 2 HgMn stars per set of targets. This programme was
part of a larger one involving other teams concerned by this
survey\footnote{074.D-0193(A), \emph{Looking deep in the CoRoT eyes}, G. Alecian
(PI), A.-M. Hubert, M. Deleuil, C. Neiner, M. Floquet, C. Martayan. }.

\section{Ground data}
\label{ground}

\subsection{VLT-FLAMES/GIRAFFE observations}
\label{giraffe}
Observations were done during the period 74A (0.5 night, 30-31 January 2005), in
the framework of the Paris Observatory GTO. In the present work, we use the data
obtained in MEDUSA mode with setup LR02 (396.4-456.7nm, R=6400). Observational
conditions were rather good with an average seeing of 0.6"-0.9". Two exposures
of 30 minutes were done for each field and an average S/N of about 70 was
reached for brightest targets. The  spectra we use in this work were obtained
after data reduction performed with the dedicated software GIRBLDRS developed at
the Geneva Observatory\footnote{http://girbldrs.sourceforge.net}. Several tasks
of the IRAF package for extraction, calibration, and sky correction of the
spectra were also used.

\subsection{Looking for HgMn stars}
\label{targets} 
About 240 stars were observed in this run, and about 60 of them were classified
in \emph{Exo-Dat} as B0-A0 (IV-V) stars. To find HgMn candidates among these 60
stars, we searched for the characteristic pattern formed by two MnII lines
(420.53nm and 460.637nm). This pattern is clearly visible in the spectra of
HD175640 (a high S/N UVES spectrum of this star was kindly communicated by S.
Hubrig, see Castelli \& Hubrig 2004), and remains well recognizable after
degrading the S/N to 100. After a close examination of the spectra of these 60
stars, one star (S1) was found to show the MnII lines pattern very similarly to
HD175640 (degraded spectrum), and two other stars (S2 and S3) were found with
the same pattern, but less clearly. The characteristics of these three stars are
shown in Table~\ref{fondpara}: their number in USNO-A2 catalog, the visual
magnitude from \emph{Exo-Dat}, and the fundamental parameters discussed in
Section~\ref{fund}.
A rough estimate of various lines strength for usual
abundances in HgMn stars was done using the COSSAM code (Stift, 2000), and we
checked the consistency of our selection of HgMn candidates by considering the
spectra at other wavelengths. These three stars were then proposed to be included
in the target allocated to the APs (Weiss, 2006).

The detailed abundance analysis of these three stars, in the next subsections,
have confirmed their identification as good HgMn candidates.

\begin{table*}
\begin{center}
\caption{Programme stars and adopted effective temperatures, surface gravities and
rotational velocities for our sample stars}
\label{fondpara}
\begin{tabular}{||c|c|c|c|c|c|c||}
\hline  \hline
Star&USNO-A2&$m_V$&\teff\ (K)&\logg\ (dex)& \vsini\ (\kms)\\  \hline
S1&0825-03036752& 12.05&13500$\pm$500&4.00$\pm$0.5&$35\pm5$	\\ 
S2&0825-03028353&13.01&12500$\pm$500&4.50$\pm$0.5&$50\pm8$	\\ 
S3&0825-02993210& 13.18&11750$\pm$500&4.00$\pm$0.5&$50\pm8$	\\ 
 \hline	
		
\end{tabular}
\end{center} 
\end{table*}

\begin{figure*}[htbp!]
\centering
\vskip1cm
\begin{tabular}{c}
\includegraphics[scale=0.5]{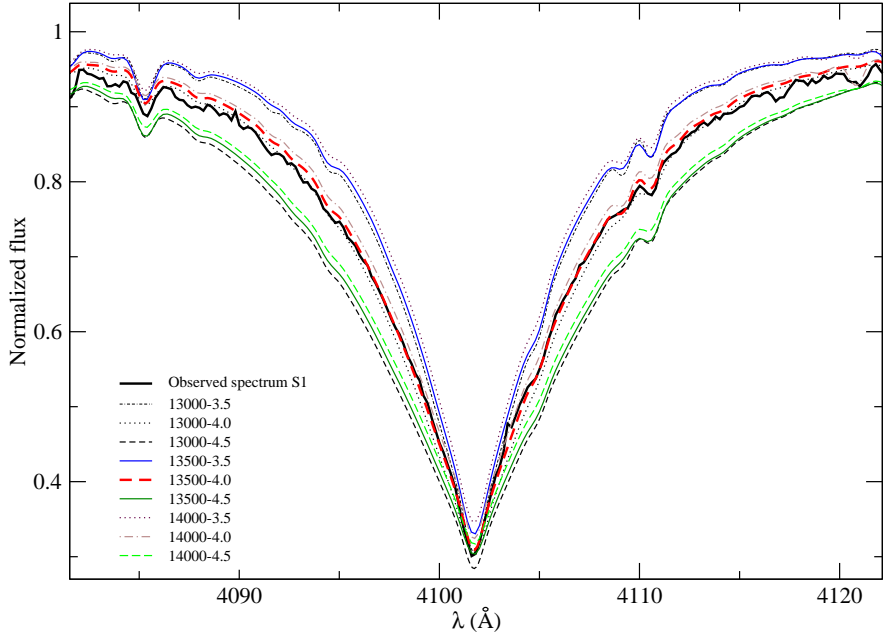}
\end{tabular} 

\caption{The effect of changing \teff\ by 500K and $\log g$\ by 0.50 dex
on the spectrum around H$_{\delta}$. The adopted temperature and gravity for S1 are 13500 K and 4.0 dex}
\label{tempX}
 
 \end{figure*}

\subsection{Fundamental parameters}
\label{fund}
The effective temperatures (\teff) and surface gravities (\logg) of the stars
were derived by adjusting a grid of theoretical Balmer lines to the observed
one. With a lack in photometric data, fundamental parameters can only be determined spectroscopically. By varying simultaneously \teff \ and \logg, we select the profile that fits best Balmer lines. LTE model atmospheres were calculated using Kurucz's ATLAS12 code (Kurucz,
2005), assuming a plane parallel geometry, a gas in hydrostatic and radiative
equilibrium and a depth-independent microturbulence. Abundances can be adjusted
individually in ATLAS12 which uses the Opacity Sampling technique for line
opacity. The increment in \teff \ and \logg \ were 500 K and 0.5 dex respectively. We have used SYNSPEC code (Hubeny \& Lanz 1992) to compute the
synthetic Balmer profiles in the wavelength range of GIRAFFE spectra
($\lambda$3950-4550 \AA). H$_{\gamma}$ and H$_{\delta}$ were generated using the
line broadening functions of Vidal et al. (1973). The solar abundances used in
SYNSPEC and ATLAS12 are those from Grevesse \& Sauval (1998). These Balmer lines are computed using a non solar ATLAS12 model. After that we select a model atmosphere, we determine the abundances of the elements and these abundances are included in the new computed model atmosphere for the same star and the final models have the abundances displayed in Table~\ref{abundances}.  \\
For stars hotter than 8000 K, Balmer lines are more sensitive to gravity than to temperature.
Figure \ref{tempX}-a shows the effects of increasing and decreasing the
effective temperature by about 500 K around 13500 K on the profile of
H$_{\delta}$. This variation of $\pm$500 K corresponds to the estimated error on
the effective temperature. In the same manner, we have tested the effects of
changing the surface gravity by about $\pm$0.50 dex which is the estimation of the error bar on $\logg$ . These tests were done for the three stars and for both H$_{\gamma}$ and H$_{\delta}$ profiles. For stars earlier than mid-A type, hydrogen lines are relatively free from metal lines and
thus an appropriate normalization to the continuum can be done. The adopted
fundamental parameters are those of the model that best fit the overall shapes
of Balmer lines. The adopted effective temperatures, surface gravities and
rotational velocities are collected in Table~\ref{fondpara} for S1, S2 and S3.

\subsection{Abundance analysis}
\label{abund}
\subsubsection{Input data}
Once the fundamental parameters were fixed, the same ATLAS12 model atmospheres
were used as an input in the synthetic spectra calculations. We have compiled a
line list in the optical domain with the most accurate atomic parameters. For
MgII, SiII, TiII, CrII, FeII, SrII and YII, we have selected lines (Table~\ref{abundetai}) from the atomic transitions list of Gebran et al.'s (2008) online table after we have checked that these lines are good enough for abundances determination in case of late B-type stars (ie. no blending and not so weak). These authors have used Kurucz's gfall.dat\footnote{http://kurucz.harvard.edu/LINELISTS/GFALL/} as an
initial line list, and they have modified the atomic parameters of the
investigated chemical elements (wavelengths, excitation potential, oscillator
strength and damping constants) by checking for more recent laboratory
determinations (as in NIST and VALD databases). For HeI, CII, PII, MnII and HgII, we
have used Kurucz's data list. The helium abundances were derived using two lines
of HeI at $\lambda$4143.8 and $\lambda$4471.5 using the broadening table of Barnar et al. (1969). We have also included data for hyperfine splitting for the selected transitions when relevant, in particular for MnII. The spectral resolution of the spectra and the effect of rotational broadening prevent us from detecting the signature of hyperfine splitting. Mercury abundance was derived using the line of HgII at 398.39 nm with the oscillator strength from Sansonetti \& Reader (2001).

\subsubsection{The method}
\label{method}
Our main interest in the abundance determination is to detect HgMn features because we cannot derive accurate chemical abundances using low resolution spectroscopy and low signal-to-noise spectra. We have derived individual chemical abundance of He, C, Mg, Si, P, Ti, Cr, Mn,
Fe, Sr and Y by iteratively adjusting synthetic spectra to the observed
normalized spectra. The synthetic spectra were computed using Takeda's (1995)
code. Takeda's procedure minimizes iteratively the chi-square statistic between
the synthetic spectrum and the observed one (See Takeda's paper for a full
description of the method).\\ 
Apparent rotational velocity (\vsini) and microturbulent velocity (\xit) were derived using the strong MgII triplet at 4480 \AA \ and a set of neighboring unblended weak and moderately strong FeII lines between 4490 and 4530 \AA \ as described in sec 3.2.1 of Gebran et al. 2008. Allowing small variations around solar abundances of Mg and Fe, we iteratively fitted these lines leaving \vsini and \xit as free parameters. The weak Fe II lines are sensitive to rotational velocity and not microturbulent velocity, the moderately strong FeII lines are sensitive to the microturbulent velocity changes. The MgII lines are sensitive to both \vsini and \xit. Each of the Mg and Fe lines yielded a set of values for $\log \epsilon$, \vsini and \xit, which were in good agreement.

Once \vsini \ and \xit \ are fixed, abundances have been derived for each line of each chemical element in our sample stars. With a resolving power of about 6500, errors on rotational velocities are more or less of the order of $\sim$15\%.  Microturbulent velocities were found to be very small, which is common in HgMn stars (see Smith \& Dworetsky 1993), thus we decided to fix them to zero for S1, S2 and S3. The found apparent rotational velocities (\vsini) are displayed in column 7 of
Table~\ref{fondpara}. \\ 

\begin{table*}
\begin{center}

\caption{Detailed abundances (in $\log{({N_X}/{N_H})}+12$ scale) for each line of each chemical element.}
\label{abundetai}
\begin{tabular}{||l|c|c|c|c|c||}
\hline  \hline
Ion&$\lambda$ (\AA)& $\log gf$ &S1 &S2 &S3 \\ \hline

HeI&4143.800&-1.19	&9.486&9.177&9.500\\
HeI&4471.500& 0.05	&9.456&9.800&9.450\\ \hline

CII&4267.261&-0.58&	7.906&7.180&7.804 \\ \hline

MgII&4390.510&-1.70&6.975&7.359&7.122 \\
MgII&4427.994& -1.20 &7.270& -&7.002 \\
MgII&4481.10&0.73, -0.57, 0.57&7.201&6.720&7.154 \\ \hline

SiII&4128.054&0.30& -&7.090&7.704 \\
SiII&4130.894&0.46&7.857&6.702&7.350 \\ \hline

PII&4420.712& -0.48 &	 7.608& -& -\\
PII&4452.472& -0.19 &	 7.420& -& -\\
PII&4466.140&  -0.28&	 7.437& -& -\\
PII&4468.000&  -0.21&	 7.760& -& -\\
PII&4475.270& 0.30 &	 7.670& -& -\\
PII&4483.693& -0.43 &	 7.502& -& -\\ \hline

TiII&4163.644&-0.13  &  -  & 6.130 &	5.880\\
TiII&4287.873&-1.79  &   - &6.206  &   5.340\\
TiII&4300.042&-0.44  &   - &6.207  &   5.246\\
TiII&4386.844& -0.96 &   - &	  -  &   5.293\\
TiII&4394.059&-1.78  &   - &6.290  &   5.450\\
TiII&4395.051&  -0.54&   - &	 -   &    5.270\\
TiII&4399.772&-1.19  &   - &6.280  &  -       \\
TiII&4417.714&-1.19  &   - &6.180  &	-     \\
TiII&4443.801&-0.72  &   - &6.170  &	 -    \\ \hline

CrII&4616.629&-1.29&	6.740&	5.510&	6.113\\ \hline

MnII&4085.390&-2.51&7.800&	-   & -	  \\	  
MnII&4252.963&-1.13&7.820&7.180 &	-   \\
MnII&4259.200&-1.58&	 -   &7.089 &7.440  \\
MnII&4292.237&-2.22&7.310&	-   &	-   \\
MnII&4275.884&-1.91&8.139&	 -  &	-   \\
MnII&4278.614&-2.51&    -  &  -     &7.080  \\
MnII&4308.158&-1.72&    -  &7.68  &6.587  \\
MnII&4310.692&-0.15&7.677&	-   &	-   \\
MnII&4325.041&-2.29&8.200&	-   &6.036  \\
MnII&4356.621&-2.02&7.809&7.246 & -	  \\
MnII&4363.255&-1.90&7.890&6.980 &  -       \\
MnII&4478.637&-0.95&8.040&	-   &6.708   \\
MnII&4503.201&-2.16&7.880&	-   &	-    \\
MnII&4518.965&-1.32&7.850&6.900 & -	   \\ \hline

FeII&4273.326& -3.25    & 7.866  &7.260 &     7.496 \\
FeII&4296.570 & -3.01 & -&7.380&-\\
FeII&4416.830&-2.60     & 7.600 & -&-\\
FeII&4491.405& -2.70    & 7.887 &7.106&      7.123 \\
FeII&4508.288& -2.21    & 7.790 &7.432 &     7.399 \\
FeII&4515.339&  -2.48   & 7.860 & 7.510&     7.349 \\
FeII&4520.224&-2.60	&-&7.150&     7.346 \\
FeII&4522.634& -2.03    & 7.760 &7.520& -  \\
FeII&4541.524&  -3.05   & 7.950 &  -& -	\\ \hline

NiII&4067.031&-1.83&	5.463& -& - \\ \hline

SrII&4215.520&-0.17&	3.850& -&	5.367\\ \hline

YII&4374.935&0.16& -&	5.440	&5.853\\ \hline

HgII&3983.90	&-1.51&7.700&	7.200&	6.065\\ \hline

\hline	
		
\end{tabular}
\end{center} 
\end{table*}

\begin{table}
\caption{Abundances (relative to the sun) table for S1, S2 and S3.}
\label{abundances}
\centering
\begin{tabular}{||c|ccc||}
\hline  \hline
Star&S1&S2&S3 \\ \hline
HeI&-1.53&-1.51&-1.17\\
$\sigma_{He}$&0.09&0.31&0.20  \\\hline

CII & -0.57&-1.30&-0.68\\
$\sigma_{C}$ &0.20&0.20&0.20 \\\hline

 MgII &-0.39&-0.50&-0.45 \\
 $\sigma_{Mg}$ &0.13&0.32&0.07 \\\hline

 SiII &0.35&-0.61&0.02 \\
 $\sigma_{Si}$ &0.20&0.19&0.18 \\\hline

 PII&2.16&-&- \\
 $\sigma_{P}$&0.12&-&- \\\hline

 TiII&-&1.23&0.43\\
 $\sigma_{Ti}$& -&0.05&0.19 \\\hline

 CrII &1.10&-0.121&0.48 \\
 $\sigma_{Cr}$ &0.20&0.20&0.20  \\\hline

 MnII &2.51&1.83&1.42 \\
 $\sigma_{Mn}$ &0.23&0.25&0.47 \\\hline

  FeII   &0.40&-0.10&-0.08\\
  $\sigma_{Fe}$   &0.14&0.16&0.13 \\\hline

  NiII   &-0.75&-&- \\
  $\sigma_{Ni}$    &0.20&-&- \\\hline

  SrII&0.92&-&2.44 \\
  $\sigma_{Sr}$&0.20&-&0.20 \\\hline

  YII&-&3.24&3.65 \\
  $\sigma_{Y}$&-&0.20&0.20\\\hline

 HgII&6.61&6.11&4.97 \\
  $\sigma_{Hg}$&0.20&0.20&0.20\\\hline

\end{tabular}
\end{table}

\begin{figure}[htbp!]
\centering
\includegraphics[scale=0.4]{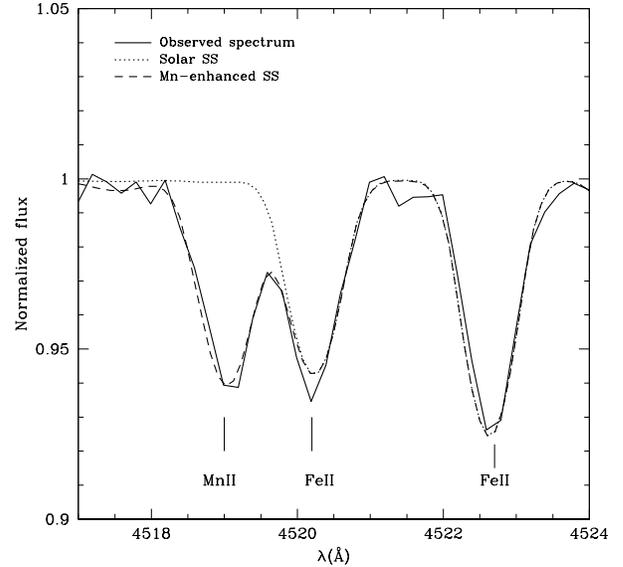}
\caption{Adjustment of synthetic spectra on the observed one for S1. The dotted
lines represent a synthetic spectra with solar abundance for all the elements.
The dashed lines represent a synthetic spectra with an enhancement of manganese
by about 300 times the solar value.}
\label{Mn-250}
 \end{figure}
 
\begin{figure*}[htbp!]
\centering
\vskip0.8cm
\includegraphics[scale=0.6]{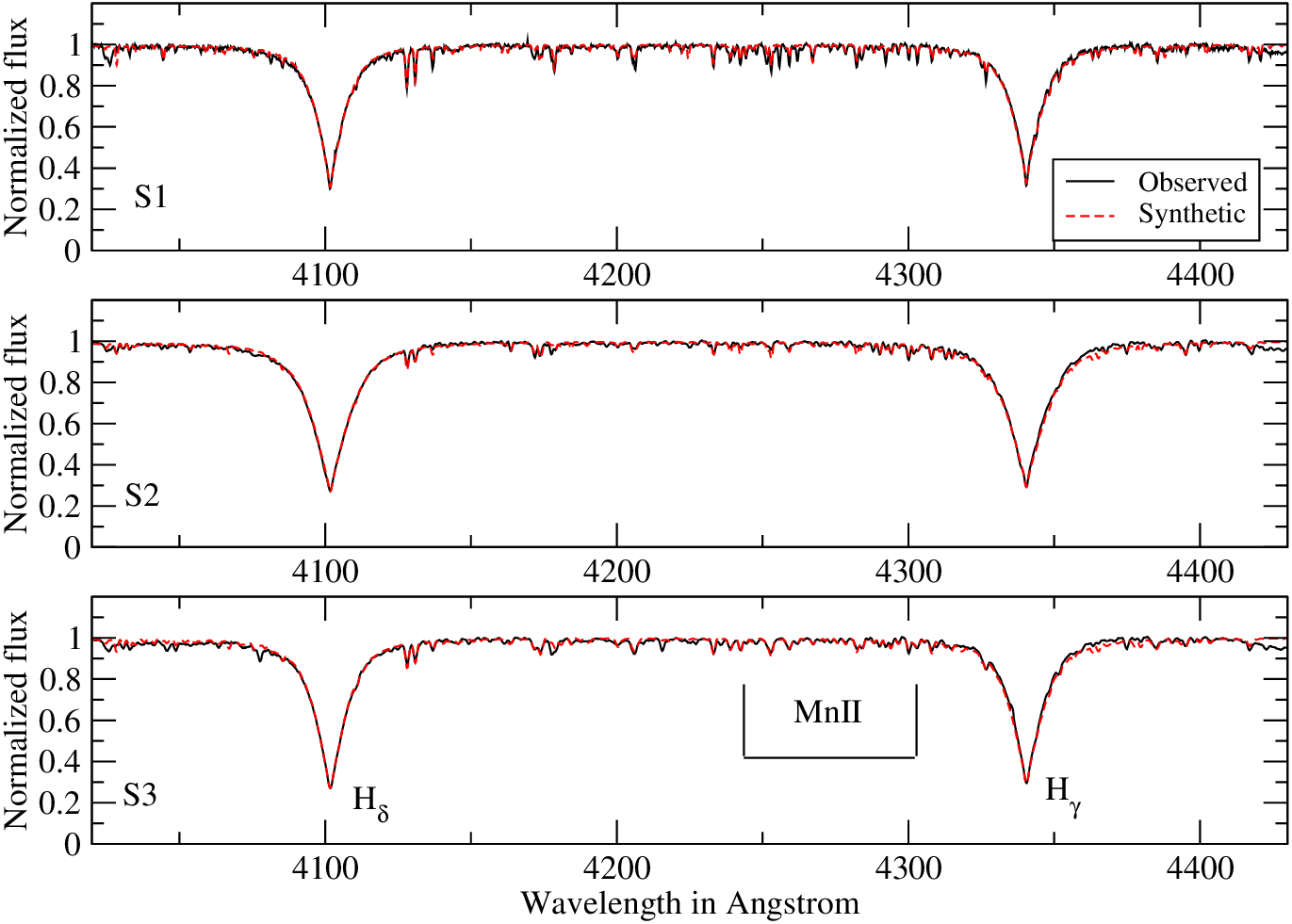}
\caption{Overplot of synthetic spectra on the observed ones for S1, S2 and S3. The range covers H$_{\gamma}$ and H$_{\delta}$ lines plus several manganese lines. The observed spectra are in solid black line and the synthetic ones in dashed red line.}
\label{fit-3stars}
 \end{figure*}

\begin{figure*}[htbp!]
\centering
\vskip1.2cm
\includegraphics[scale=0.6]{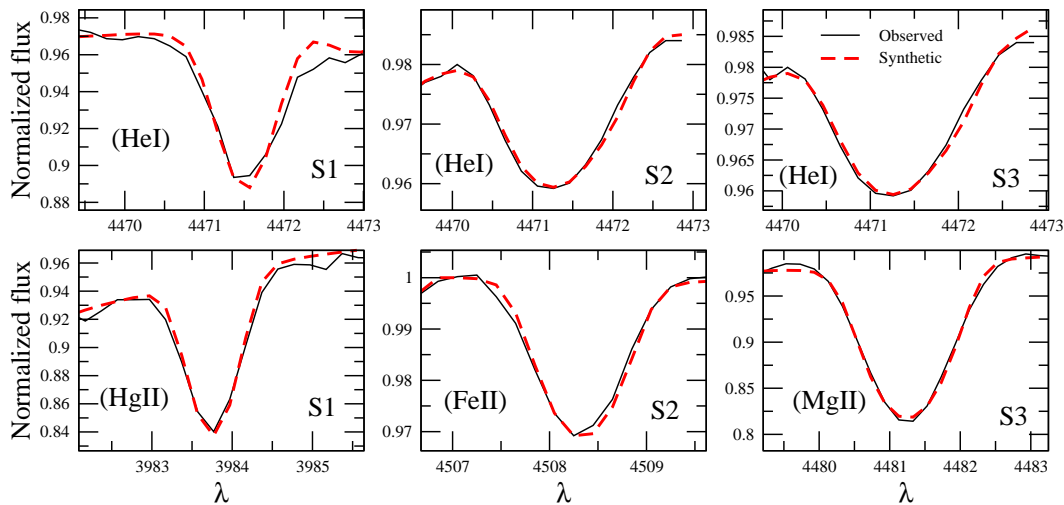}
\caption{Several examples of line fitting using different chemical species for the three stars. Helium, mercury, iron and magnesium lines are displayed. The observed spectra are in solid black line and the synthetic ones in dashed red line.}
\label{6fits}
 \end{figure*}

\begin{figure}[htbp!]
\centering
\includegraphics[scale=0.4]{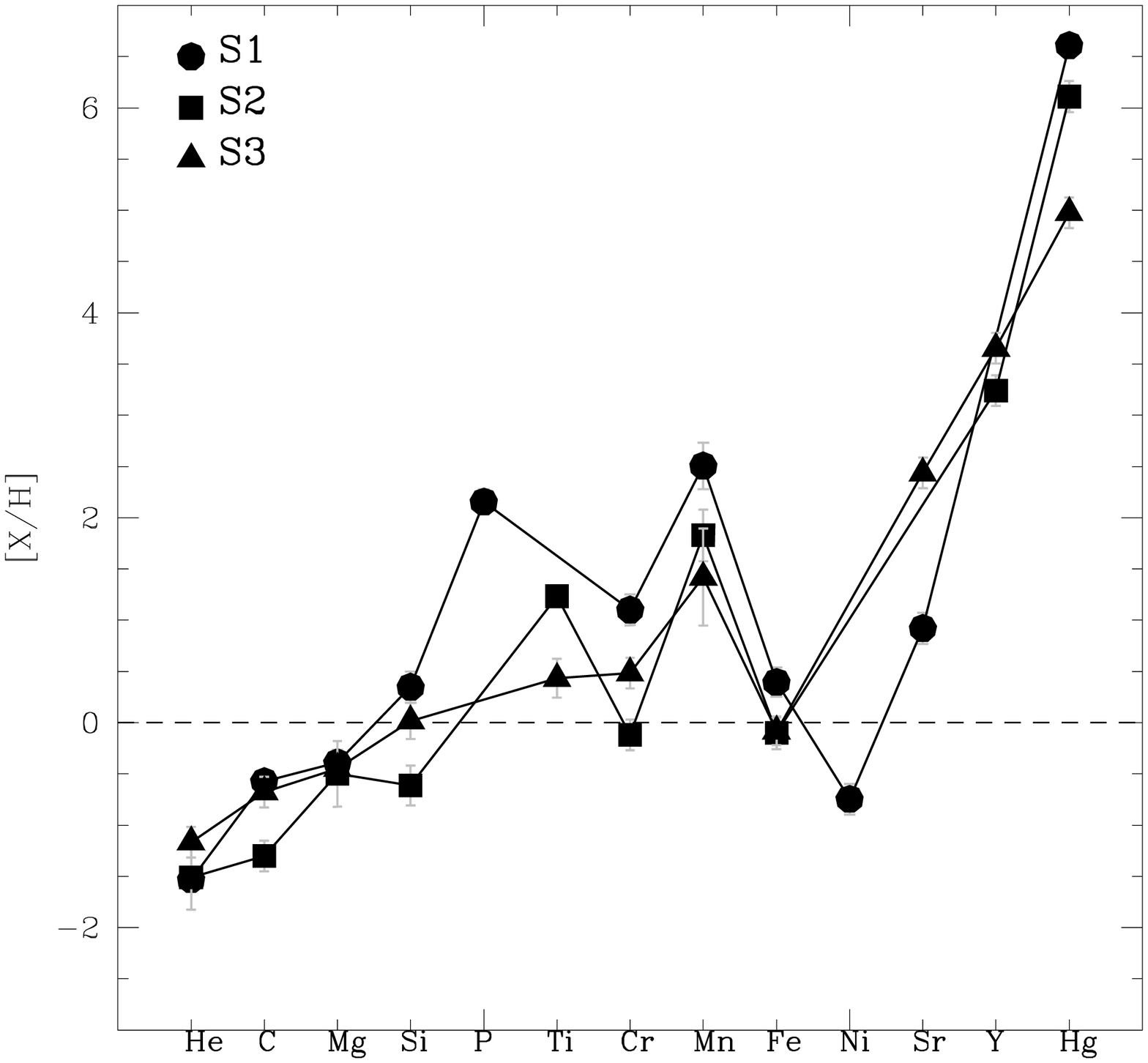}
\caption{Abundances pattern for our sample stars}
\label{pattern}
 \end{figure}

The derived chemical abundances relative to the sun\footnote{($[\frac{X}{H}]=\log(\frac{X}{H})_{\star}-\log(\frac{X}{H})_{\odot}
$)} for each lines used for this analysis are shown in Table \ref{abundetai}. The final abundances for each star, and their respective errors are displayed in Table~\ref{abundances}. 
The abundance [X/H] does not include the rescaling of the H abundance in order to account for the He deficiency.
We have assumed that the abundances derived from several lines for a chemical element, follow a Gaussian distribution. The errors on the element abundances are thus standard deviations. If the abundance is derived from only one line, we assign to it an error of 0.20 dex. We have also tested the effect of changing the atmospheric parameters on the derived abundance. Changing the effective temperature by an amount of $\pm$ 500 K affect the mean derived abundances by an amount of $\sim$0.1-0.3 dex depending on the element and of course the HgMn trends are not affected. Figure \ref{Mn-250} displays an example of the iterative fitting of the MnII line at $\sim$4519 \AA \ for S1. The synthetic spectrum with solar abundances is in dotted line and the manganese
enhanced spectrum is in dashed line. In case of S1, manganese is about 300 times
higher than solar. We have displayed in Fig.\ref{fit-3stars} the overall fit of the three star between 4000 and 4400 \AA. In this region, we have the two Balmer lines ($H_{\gamma}$ and $H_{\delta}$) and a large number of MnII lines. In Fig.\ref{6fits}, we show some examples of line fitting for S1, S2 and S3 and for several chemical elements as He, Hg, Fe and Mg. Due to the low signal-to-noise, some of the weak lines couldn't be analyzed as the titanium in S1 or nickel in S2 and S3. It is clear that all these 3 late-type B stars present HgMn features. Large overabundances are found for P, Mn, Sr, Y and Hg and underabundances in He and Ni. It also appears that the hotter HgMn star have the stronger helium deficiency and manganese enhancement. Figure \ref{pattern} displays the abundances pattern of S1, S2 and S3. One should be aware that the present analysis is based on low resolution and low signal-to-noise spectra, but the important result is that the HgMn pattern is clearly displayed in all three stars.

\section{CoRoT Observations}
\label{corotobs}
Stars S1, S2 and S3 were observed by CoRoT during the
first \emph{long run} towards the galactic anticenter (LRa01). This
\emph{long run} lasted from the 24th of October 2007 to the 3rd of March 2008,
i.e. about 130 days. All the results presented here are based on the \emph{N2
data} provided by the CoRoT data
center\footnote{http://idoc-corot.ias.u-psud.fr/index.jsp}, released on October
29th, 2008. We have proceeded to the additional treatments of the data (see
Sec.\ref{dataprep}) using IgorPro~v6 (from
\emph{Wavemetrics})\footnote{http://www.wavemetrics.com/}, and
\emph{Period04}~v1.0.4 (Lenz \& Breger, 2005) for the analysis of frequencies.

In this section, for the sake of conciseness, we will present in details the
treatment of data for the star S1 only.

\subsection{Data preparation}
\label{dataprep}
A detailed presentation of the CoRoT data is given by Auvergne et al. 2009. For
each of our three stars, CoRoT provided chromatic lightcurves, in the
oversampled mode (32 sec.)\footnote{Chromatic and oversampled data are available
only for a limited number of targets of the exoplanet channel.}. 
These chromatic lightcurves are not provided to give color informations on targets, because the separation between the colors (which is realised by the ground based data processing) is arbitrary. The color separation was implemented to help in transit detection of exoplanets, but does not give workable information on star's properties, at least at the present stage of data treatment. 
The colored fluxes are strongly sensitive to depointing fluctuations and
consequently the jitter corrections on colors are not very efficient. The white
light flux is less sensitive to jitter because on the aperture edges the PSF
slope is small. So it is recommended to workÊ
first for detection on white lightcurves.

The original set
of \emph{N2} lightcurves (red, green, and blue) for S1 consists in 338729 points
of measurement (electrons vs. CoRoT-Julian days), corrected for main
instrumental effects. We have first selected the points flagged as "valid"
($status=0$) by the \emph{N2} pipe-line. Other points are generally affected by
the so-called SAA (South Atlantic Anomaly) and/or  other instrumental effects.
However, some abnormal points survive this selection and need to be suppressed
manually. At the end of this selection, one has 300029 points of measurement
(over 131 days). Most of them (98\%) are spaced by 32 sec, but this whole
selection process introduces some gaps, often smaller than about 1100 sec due to
SAA. The SAA is crossed by the satellite 8 time per day, and lasts about 10 mn. There are also gaps due to various interruptions (normal and abnormal) of data acquisition. The largest gap is 13.7 hours from JD=2861.23 due to missing data. The next
step consisted in correcting the monotonic long-term variation of each curve by
a parabolic fitting. This variation is partly due to a pointing shift, and
partly due to instrumental drift. After this correction we ended with three
lightcurves summarized by the following values (average level, standard
deviation): Red (391.373 kE, $\sigma=1.293$ kE), Green (107.826 kE,
$\sigma=0.435$ kE), Blue (242.617 kE, $\sigma=1.056$ kE). Finally, for analysis
of frequencies (Sect.~\ref{dataanal}), we have built the \emph{white} lightcurve
 by simply
summing the three chromatic lightcurves, and we achieved a binning of 512 sec to
reduce the noise (hereafter, all the presented plots and results are based on
data with a binning of 512 sec). A similar treatment was done for stars S2 and
S3.

\subsection{Data analysis and results}
\label{dataanal}

\subsubsection{Star S1}
\label{S1}

\begin{figure*}
\centering
\includegraphics[width=18cm]{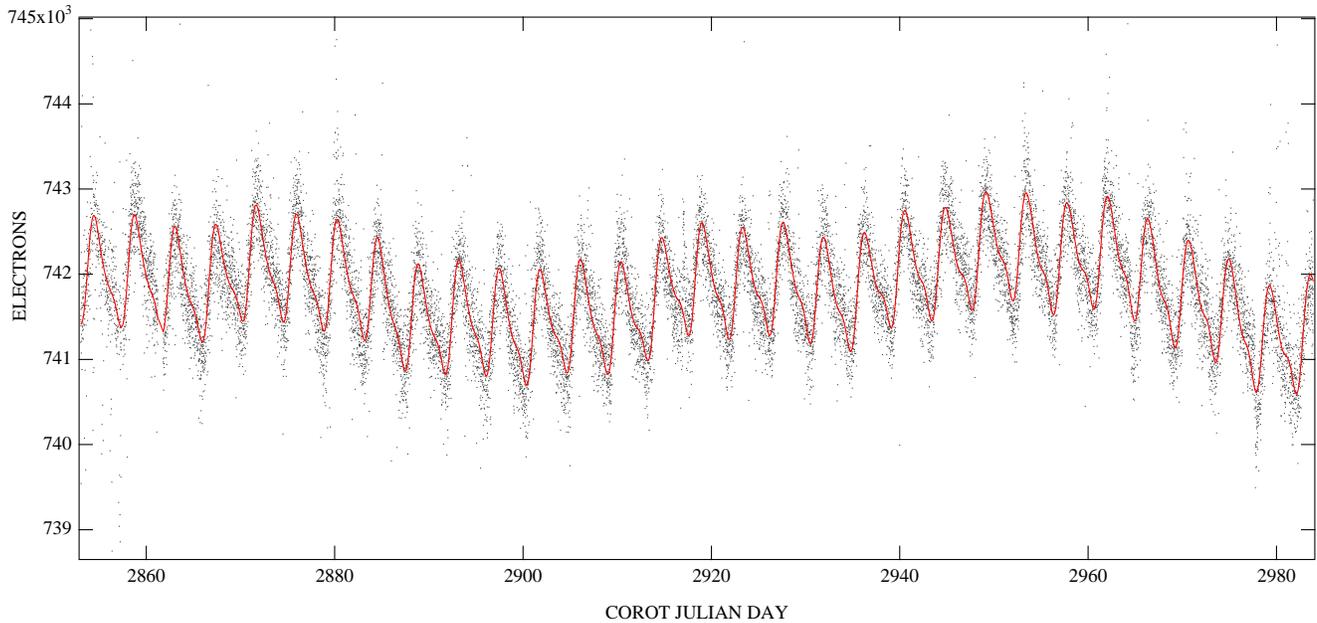}
\caption{
S1 \emph{white} lightcurve and the fit obtained using the frequencies of
Table\ref{freqS1} (ELECTRONS vs. CoRoT JD).
}
\label{fig_white_fit}
\end{figure*}

\begin{figure}
\includegraphics[width=9cm]{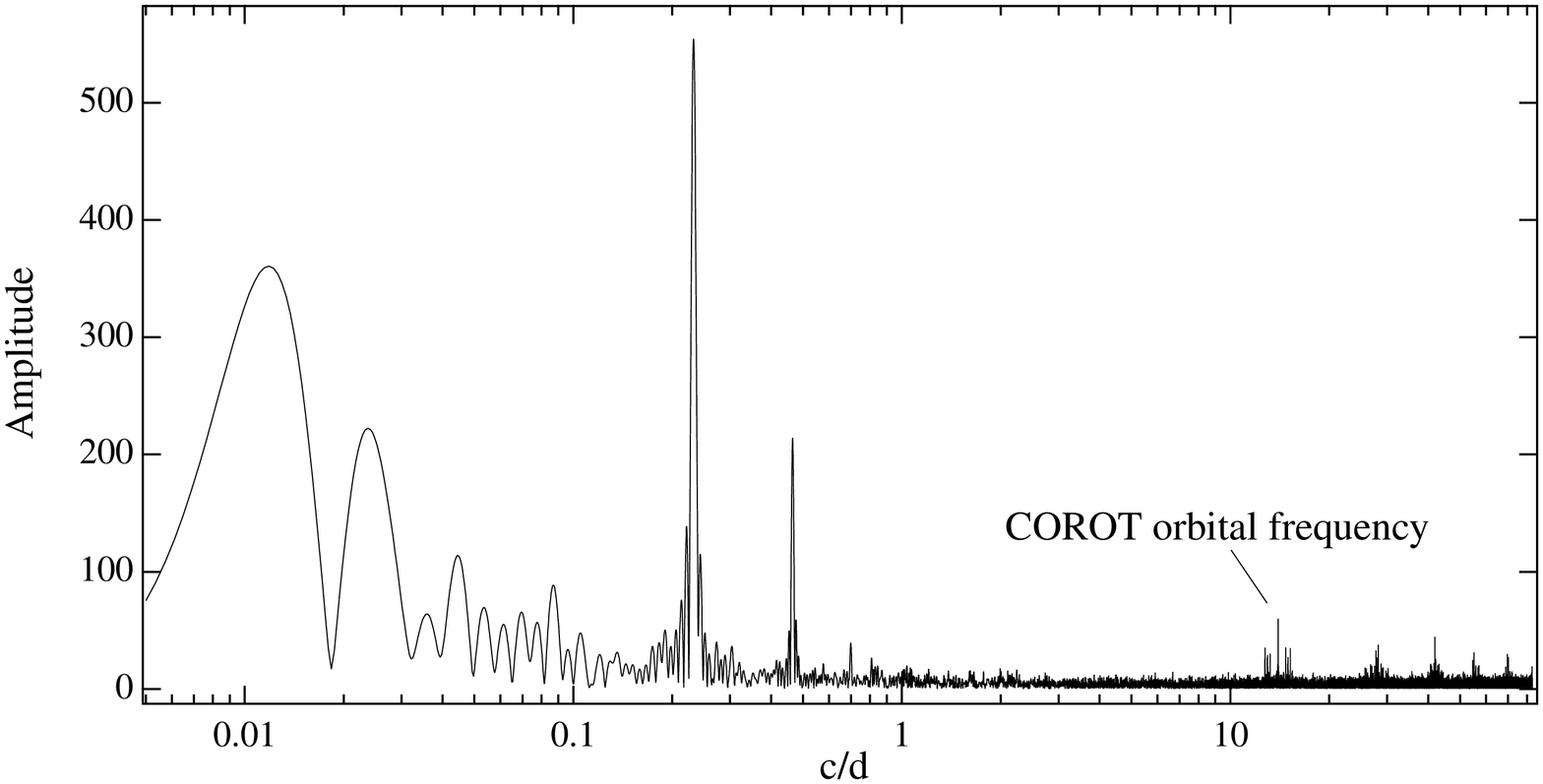}
\caption{
Full S1 spectrum (amplitude vs. cycles per day) calculate with Period04 from the
\emph{white} lightcurve.
}
\label{fig_Spect_S1_full}
\end{figure}

\begin{figure}
\includegraphics[width=9cm]{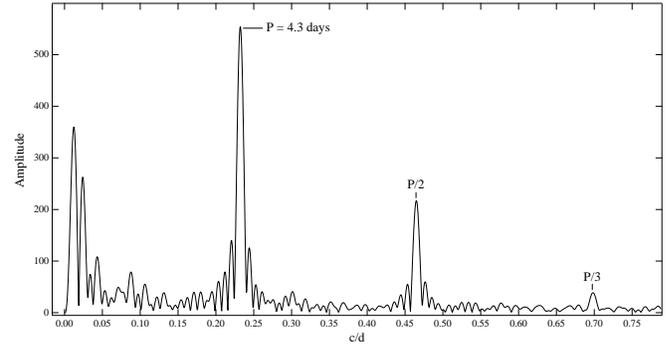}
\caption{
Details of S1 spectrum. Zoom in Fig.\ref{fig_Spect_S1_full}, with a linear
scale.
%, around the period of 4.3 days and its harmonics.
}
\label{fig_sect_zoom_S1}
\end{figure}

The \emph{white} lightcurve (points) is shown in Fig.\ref{fig_white_fit}. The
solid line is the fit obtained with Period04, using the frequencies listed in
Table~\ref{freqS1}. This list of frequencies corresponds to frequencies
with the highest amplitudes, from which the well-known CoRoT-orbital ones and its
harmonics were removed. The full spectrum is shown in
Fig.\ref{fig_Spect_S1_full}. The CoRoT orbital frequency at 13.97 c/d
(161.7$\mu$Hz) and its harmonics are clearly visible. The low frequencies around
0.01 c/d are related to the long-term variation  noticeable in
Fig.\ref{fig_white_fit}. We cannot interpret safely these long-term
variations which have periods comparable to the run duration. They do not
correspond to known instrumental effect.
Fig.~\ref{fig_sect_zoom_S1} presents the detailed spectrum around 0.23 c/d. The
feature around the peaks are due to the spectral window. In our opinion, the
period $P\approx 4.3$ days, with two harmonics ($P/2$, $P/3$) is an intrinsic
variation of star S1 and its amplitude ($\Delta{m}/m$) can be estimated
to be about 1.6 mmag. There is not any known CoRoT-periods around $P$, $P/2$
and $P/3$. It is easily visible in Fig.\ref{fig_white_fit} (the ``high"
frequency variation).

\begin{table}
\caption{List of frequencies for star S1 (from Period04), used to calculate the
fit of Fig.\ref{fig_white_fit}}
\label{freqS1}
%\centering
\begin{tabular}{||l|c|c|c|c|c||}
\hline  \hline
N&c/d&$\mu$Hz&Ampl.&Phase&Period (days)\\  \hline
1&0.0114&0.132&413.9&0.7943&88.007 \\
2&0.0146&0.169&106.3&0.8962&68.412 \\
3&0.0251&0.290&174.6&0.0849&39.880 \\
4&0.0436&0.505&74.1&0.7371&22.913 \\
5&0.0877&1.015&73.5&0.4033&11.401 \\
\textbf{6 (P)}&0.2324&2.689&552.9&0.9211&4.304 \\
\textbf{7 (P/2)}&0.4648&5.379&212.7&0.6286&2.152 \\
\textbf{8 (P/3)}&0.6982&8.081&37.1&0.5796&1.432 \\ \hline

\end{tabular}
\end{table}

\subsubsection{Star S2}
\label{S2}

\begin{figure}
\includegraphics[width=9cm]{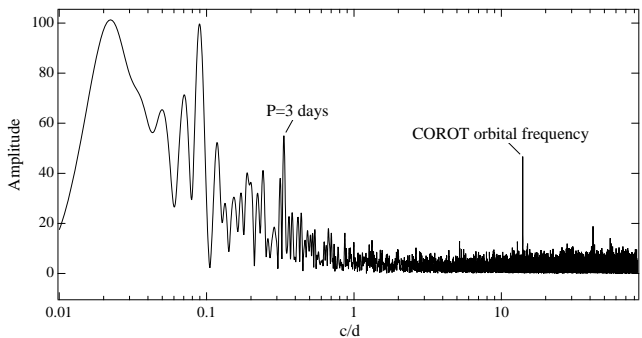}
\caption{
Full S2 spectrum (amplitude vs. cycles per day) calculated with Period04 from
the \emph{white} lightcurve.
}
\label{fig_Spect_S2_full}
\end{figure}

The lightcurves of star S2 are much noisier than those of star S1. In addition,
the blue lightcurve presents some strong discontinuities and perturbations from
the date JD~$\approx 2931.36$. We have then selected only points between
JD~$=2853.036$ and $2931.35$. These data lead to the rather noisy spectrum shown
in Fig.~\ref{fig_Spect_S2_full}. Again, the CoRoT orbital frequency (and
harmonics) dominates at 13.97 c/d (and above). The only other significant peak
is for $0.337 \pm 0.01$ c/d, very close to a period of 3 days, which may be
related to the 1 day period often present in CoRoT data. We cannot find
confidently from these rather poor data frequencies intrinsic to the star S2.

\subsubsection{Star S3}
\label{S3}

\begin{figure}
\includegraphics[width=9cm]{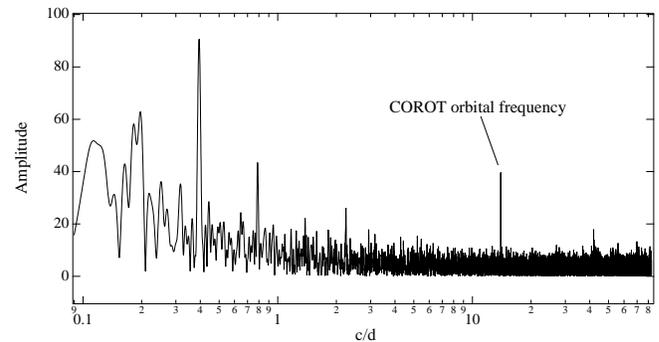}
\caption{
Full S3 spectrum (amplitude vs. cycles per day) calculated with Period04 from the
\emph{white} lightcurve.
}
\label{fig_Spect_S3_full}
\end{figure}

\begin{figure}
\includegraphics[width=9cm]{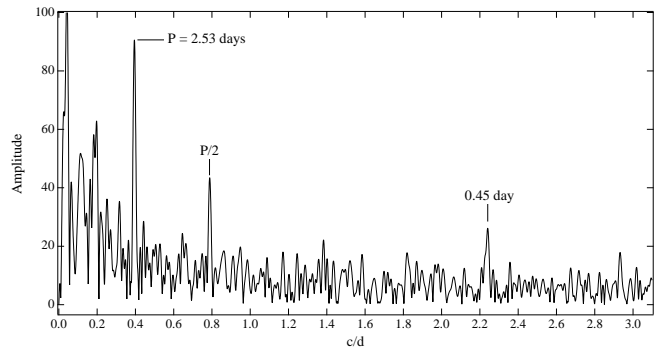}
\caption{
Details of S3 spectrum. Zoom in Fig.\ref{fig_Spect_S3_full}, with a linear
scale.
%, around the period of 2.53 days and its harmonic.
}
\label{fig_sect_zoom_S3}
\end{figure}

As in the previous case, lightcurves of star S3 present several discontinuities 
and perturbations, but not exactly at the same dates. However, compared to star
S2, the lightcurves are less noisy and present interesting features. We have
considered the data between JD~$=2853.036$ and $2925.933$ (73 days) and we
show the corresponding full spectrum in Fig.~\ref{fig_Spect_S3_full}, the
details at low frequency range ($<3$ c/d, linear scale) are in
Fig.~\ref{fig_sect_zoom_S3}. There are clearly two peaks above the noise, the
first one at $0.395\pm 0.01$ c/d (P $\approx 2.53$ days) and the second one  at
$0.789\pm 0.01$ c/d which is possibly the first harmonic (P/2). 
There is a peak around 0.19 c/d, but as one can see in better detail in Fig.~\ref{fig_Spect_S3_full}, it does not emerge from the surrounding patterns as clearly as the previous peaks.
Another isolated
peak emerges around $2.24$ c/d (0.446 day), but it seems less significant than
the ones at $0.395$ and $0.789$.

\section{Theoretical models for HgMn stars}
\label{TheoHgMn}

The models for this paper are calculated with atomic diffusion that is known
from first principles. In radiative zones this implies taking into account
atomic diffusion including gravitational settling, thermal diffusion and
radiative accelerations, in addition to the purely diffusive term. The
interactions between different diffusing species are also taken into account.
Convection and semi-convection are modeled as diffusion processes as described
in Richer et al. (2000) and Richard et al. (2001). The detailed treatment of
atomic diffusion is described in Turcotte et al. (1998). In these models the
Rosseland opacity and radiative accelerations are computed at each time step in
each layer for the exact local chemical composition using monochromatic
opacities of the OPAL group Iglesias \& Rogers (1996). The radiative
acceleration for each element is from Richer et al. (1998) with correction for
redistribution of momentum to electrons and others ions from Gonzalez et al.
(1995) and LeBlanc et al. (2000). All models have an initially homogeneous solar
composition.

For this preliminary study we include mixing in all the models in an ad-hoc
parametric way as in Richer et al. (2000) where the parameters specifying
effective transport coefficients are defined. All our models were computed with
a effective transport coefficient, $D_{\rm T}$, 50 times larger than the He
atomic diffusion coefficient at $\log T = 5.3$ and varying as $\rho^{-3}$ deeper
in the star (for temperature below $\log T = 5.3$ we assume a complete mixing).
The choice of the layer with $\log T =5.3$ as lower boundary of the mixed zone is justified by the fact that the iron accumulation (due to diffusion process) around that layers, leads to the appearance of a convection zone. The existence of HgMn phenomenon shows that full mixing cannot extend deeper, otherwise superficial abundance anomalies would not appear. According to the diffusion model, this lack of deeper mixing may be due to the slow rotation of HgMn stars. To better model HgMn stars one would like to compute atomic diffusion in higher layers (with lower temperature, including the atmosphere). But the monochromatic opacities available to us are not sufficiently accurate to compute reliable radiative accelerations at low temperature (LeBlanc et al., 2000; Richer et al., 1998) and also, moving the boundary of the fully mixed zone to lower temperature leads to numerical difficulties in the present version of our code. 
Therefore, in our models, mixing impedes atomic diffusion in layers with $\log T < 5.3$. Notice that we have not adjusted the mixing depth to fit the surface abundances of our observed stars, because these anomalies are expected to be due to atomic diffusion in the atmosphere.
For layers deeper in the star, where $\log T \geq 5.3$,
the effective transport coefficient is adjusted to rapidly decrease the efficiency of the mixing so in these layers the effect of atomic diffusion leads to chemical stratification and to an accumulation of iron which affect the opacities and the pulsational properties of the models.

\begin{figure}
\includegraphics[width=8.5cm]{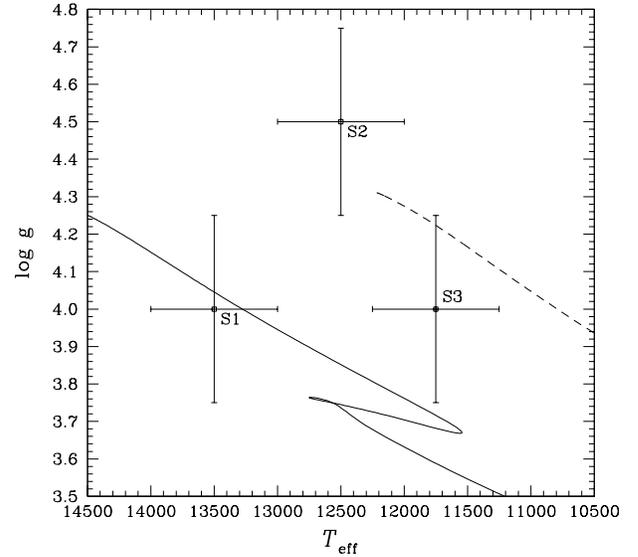}
\caption{Evolution of models of 3.0$M_{\odot}$ and 4.0$M_{\odot}$ in the $\log g
- T_{\rm eff}$ plan. S1, S2 and S3 are shown with error bars.}
\label{TeffLogg}
\end{figure}

\begin{figure}
\includegraphics[width=8cm]{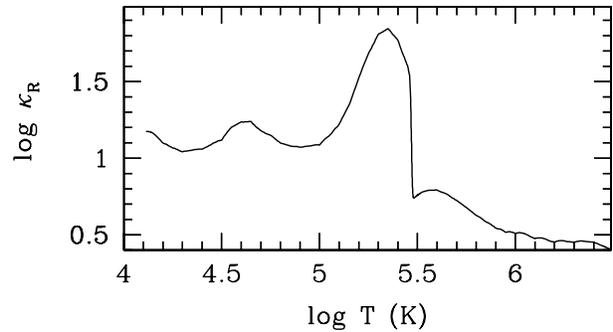}
\caption{Rosseland mean opacity versus temperature in 
    the 4.0$M_{\odot}$ model at the age of 100~Myr.}
\label{kappa}
\end{figure}

\begin{figure}
\includegraphics[width=8cm]{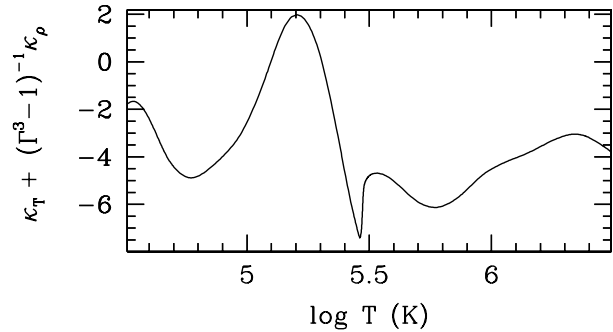}
\caption{Logarithmic opacity derivatives vs temperature for the 4.0$M_{\odot}$ 
model at the same age as in Figure~\ref{kappa}.}
\label{dkappa}
\end{figure}

We have computed models of 3.0$M_{\odot}$ and 4.0$M_{\odot}$ which are in the
mass range of S1 and S3 stars (see Figure~\ref{TeffLogg}). As shown in
Figure~\ref{kappa} these models have a important bump in the opacity profile
around $\log T = 5.3$, at this temperature radiative accelerations lead to an
iron overabundance. In this region, see Figure~\ref{dkappa}, oscillation could
be driven by $\kappa$-mechanism as
${d\over dr}(\kappa_T + {\kappa_\rho\over\Gamma_3-1})>0$, where 
$\kappa_T=(\partial\ln\kappa/\partial\ln T)_\rho$ and 
$\kappa_\rho=(\partial\ln\kappa/\partial\ln \rho)_T$.

The seismic properties of the models were studied with the help of the {\sl
nadrot} code (Dziembowski, 1977). Both the 3.0$M_{\odot}$ and the 4.0$M_{\odot}$
models show positive growth rates for non radial gravity modes, and all p-modes
are stable. In the 4.0$M_{\odot}$ model, the periods can be as short as 16 hours
and as long as 2 days for young models, whereas no mode under 33 hours is
overstable at the end of the main-sequence. The upper limit for the period, when
the 4.0$M_{\odot}$ leave the main-sequence, is hard to define as the number of
modes identified by the pulsation code is limited, but we find modes with period
as long as 4.4 days to be overstable while modes with period above 6.5 days are
stable. The 3.0$M_{\odot}$ model have a similar range in periods for young
models but show a smaller shift to longer periods during evolution with no
overstable modes with period longer than 2.5 days until all modes become stable
around 250~Myr.

\section{Discussion}
\label{discuss}

The variations we find for stars S1 and S3 can be reasonably considered as
intrinsic to these stars.

The VLT/GIRAFFE spectra used in this work are medium resolution spectra, then the v.sini (Table~\ref{fondpara}) we obtained are not very accurate and can be rather considered as upper limits of the values one could expect to determine from higher resolution spectra. Since v.sini fixes a lower limit of equatorial velocity, it gives an indication about the upper limit of rotational period. The upper limits of rotational period which can be inferred from Table~\ref{fondpara}, appear to be in the same ranges as the periods we find for S1 and S3. Therefore, one cannot exclude at this stage that the variations found in our lightcurves are due to rotational modulations (see for instance, Lanza et al. 2008).

\subsection{Rotational modulation hypothesis}
\label{rothyp}
According to the discussion by Clarke (2003), surface inhomogeneities like spots
or patches can produce photometric variations, but only the
fundamental and the second harmonic could be present and not the third harmonic
as we observe in S1. However, it is difficult to assert that high degree harmonics cannot be found in Fourier analysis in case of complex superficial structures. Magnetic Ap stars are known to have large scale and organized strong magnetic fields, with abundance patches of many metals. These anomalies are qualitatively well understood in the framework of atomic diffusion but the detailed quantitative modelling of that process is difficult, and sometimes strongly suggests that diffusion has to work with other processes like inhomogeneous winds (Babel 1992). One notes also that a complete discussion should also consider light-induced drift and ambipolar diffusion. Such superficial structures have been studied from several decades from line shape variations (for recent studies, see for instance observations by Kochukhov et al. 2004, and modelling by Alecian \& Stift 2007), and one would not be surprised if such complex structures cause photometric variations (in addition to those due to the roAp phenomenon for cool magnetic Ap stars). If our stars were not HgMn, but magnetic Ap stars, one would consider attentively rotational modulations to explain lightcurve variations.

Considering that the 3 stars discussed in this work are likely to belong to the HgMn group, is the hypothesis of abundance patches or spots relevant in explaining the photometric variations we observe? Observational facts are that there is not yet any confirmed detection of magnetic fields (Shorlin et al. 2002, Wade et
al. 2006)  nor any detection of photometric variations (Adelman et al. 2002) in HgMn stars from ground based observations. There are some possible cases of few hundred Gauss detections (Hubrig et al 2006a), but they need to be confirmed, and they are not typical for the HgMn group.
Following these constraints, standard diffusion modelling of HgMn stars consists in estimating the diffusion velocities and the resulting vertical stratifications of elements (equilibrium approximation), assuming stable atmospheres without magnetic fields. This modelling produced encouraging results by explaining main trends of metal overabundances in these stars (Smith \& Dworetsky, 1993), but cannot yet reproduce the large variety of detailed observations of individual stars. An important reason of the difficulty in modelling individual stars is that diffusion is a slow\footnote{Diffusion time scales in atmospheres extend from some years to centuries according to the considered element.} (and then fragile) time-dependent process which, at a given time, has built up abundance stratifications which are different from those computed assuming equilibrium (the finale stage of an ideal unperturbed time-dependent process). 

Because, in the standard model, atomic diffusion is supposed to work alone, any perturbation, as it could exist in a real HgMn star, can modify the picture of the stratification build-up process. For instance, effects induced by binarity, slow matter circulation, weak turbulence or wind can exist in HgMn stars. If so, they could be undetectable but strong enough to affect the stratification process. 
Among possible theoretical scenarios, one can consider the existence of weak, unorganized magnetic fields (hardly detectable). In magnetic Ap stars, where magnetic fields strengths can be larger than several kG, the component of ion's diffusion velocities orthogonal to the field lines is significantly reduced above $\log \tau \approx 0$ and vanishes as matter density decreases. This leads to the formation of non-uniform (horizontally) clouds of metals, anchored in the usual line-forming region (just above $\log \tau \approx 0$). In HgMn stars, if weak fields exist, the effect on diffusion velocities will be the same as in magnetic stars, but much higher in the atmosphere (for instance, above $\log \tau \approx -4.0$ for fields $\le 1$ kG, see Alecian \& Stift 2006). This is due to the fact that the diffusion velocity of an ion in a magnetic field is slowed down by a factor which varies more or less as the square of the proton number density (see the detailed formulae in Sec. 4.1 of Alecian \& Stift 2006). So, a weak magnetic field which will have no noticeable effect in usual line-forming region could, however, force elements to remain bound to the star. In that case, one could expect that there may be high altitude additional clouds for some elements, forming abundance patches (related to the magnetic field orientation, possibly unorganized) and superimposed to the uniform stratifications predicted by the standard diffusion models.

Recent works (Adelman et al. 2002, Kochukhov et al. 2005, Hubrig et al., 2006b) have detected in some HgMn stars variations in some sparse line profiles and interpreted them in terms of abundance spots. As discussed just before, these observations are not necessarily in contradiction with the diffusion model. Moreover, one can note that these variations of line profiles are essentially found for elements having low cosmic abundances (Hg, Y, Pt, Zr), which bring an additional support to the diffusion model. Indeed, according to what one knows about radiative accelerations, these kind of elements are generally strongly pushed up by the radiation field\footnote{Because photoabsorption by bound-bound transitions remain unsaturated until huge overabundances are obtained.} and can more easily accumulate than abundant elements in the upper atmosphere (in addition to deeper stratifications), where a weak magnetic field becomes able to impede atomic diffusion to expel that element toward the circumstellar medium.

According to the published profile variations mentioned above (papers do not give any estimation of the photometric variations that can be induced by these profile variations), one can hardly consider that, even if our stars would have same kind of line profile variations, they could produce the 1.6 mmag photometric variations we observed in CoRoT lightcurves. Remember that the CoRoT lightcurves are integrated photon flux from about 400nm to 1000nm. To have regular photometric variations (as the ones we observed) from abundance spots, equivalent width of lines should have to vary more or less in phase. This is not observed and this is not consistent with abundance patches which will be distributed differently according to elements. It does not mean that line profile variations could not coexist with photometric variations. We just say that these two types of variations are related to two different mechanisms.

Our conclusion is that the variations observed recently in some sparse line profiles for some HgMn stars remain consistent with the diffusion model, but could hardly explain the photometric variations we observed in CoRoT lightcurves. 
Nevertheless, even if our feeling
is that these variations are not rotational modulations, more precise measurements of v.sini are clearly needed to discriminate pulsations and rotational modulations, and check whether these stars are multiple or not. This is planned for the near future.

\subsection{Pulsation hypothesis}
\label{pulshyp}
In the hypothesis of pulsations, we have a significant 4.3 days period in S1,
and 2.53 days period for S3. They are compatible with periods of non-radial
g-modes predicted by models presented in Sec.\ref{TheoHgMn}. According to these
models and the study of Turcotte \& Richard (2003), these pulsations should be
of the same type as those of SPB stars (Dziembowski et al. 1993). However, one
cannot consider these stars as belonging to the SPB class, since
stars of this latter class have more or less solar abundances. In
average SPB stars  are hotter than HgMn stars, and have higher gravity. But, the
boundary at the cool edge and at the low gravity edge of SPB's (De Cat, 2007)
overlaps respectively the hot/large ones of HgMn group. In the same way, the
periods we find for S1 and S3 are little bit longer and with smaller amplitudes
(less than or around 1.6 mmag) than what is generally found for SPB's, but here
again these variations are compatible with those observed for some modes of
SPB's. If pulsations were confirmed in the future in HgMn stars, differences
could come from the lack of superficial convection zone which leads to strong
metal stratifications in outer layers in HgMn stars(including the atmospheres).
One can speculate that could cause stronger damping of those modes compared to
SPB stars.

\section{Conclusions}
In this work, we first identified from multiobject spectroscopy with moderate
resolution (using VLT-GIRAFFE spectrograph), 3 faint stars presenting the
characteristic features of HgMn types, and eligible for observation with the
exo-CCD's of the CoRoT satellite. The analysis of their lightcurves, obtained
during the run LRa01 of CoRoT, shows periodic variations for at least two of
theses stars. The periods of these variations are compatible with theoretical
predictions. This could be the first detection of pulsations in HgMn stars.
However, further ground based high resolution spectroscopic observations are
needed to confirm the HgMn nature of these stars and that these variations are
pulsations and not rotational modulations.

\begin{acknowledgements}
      The ground-based spectra used in this work are from a multi-purpose survey
      to prepare several CoRoT programmes. We would like to thank C.~Martayan
      who realized the observations corresponding to the anticenter direction
      with (ESO) VLT/FLAMES-GIRAFFE spectrograph, for the data reduction and for
      his help during preparation of the survey. Thanks are also due to
      the CoRoT exoplanet team, and specially M. Deleuil, C. Moutou, and J.C.
      Meunier to have make available for us the necessary data of Exo-Dat
      Catalog prior its official release. We would like also to thank CNES (the
      french space agency) for a financial contribution. \end{acknowledgements}

%\appendix{}
%
%\section{Evaluation of the effect of a radial velocity on the flux}
%\label{evalflux}


\begin{thebibliography}{}

\bibitem[]{} {Adelman}, S.~J., {Gulliver}, A.~F., {Kochukhov}, O.~P., \&   
	{Ryabchikova}, T.~A., 2002, ApJ, 575, 449
\bibitem[1981]{AlecianMi81} Alecian, G., \&  Michaud, G., 1981, ApJ, 245, 226

\bibitem[2006]{als06} Alecian,~G., \& Stift,~M.J.  2006, A\&A, 454, 571

\bibitem[2007]{AlecianSt2007} Alecian, G., \&  Stift, M.~J., 2007, A\&A, 475, 659

\bibitem[Auer, \&  Mihalas(1973)]{1973ApJS...25..433A} Auer, L.~H., \& Mihalas, D.\ 1973, \apjs, 25, 433

\bibitem[Auvergne et al. (2009)]{} Auvergne, M. et al., 2009, A\&A, in press

\bibitem[1992]{babel92} Babel,~J., 1992, A\&A, 258, 449
\bibitem[Baglin2006]{Baglin2006} Baglin, A. 2006, The CoRoT mission, pre-launch status, stellar seismology and planet finding (M.Fridlund, A.Baglin, J.Lochard and L.Conroy eds, ESA SP-1306, ESA Publication Division, Noordwijk, The Netherlands).

\bibitem[Barnard et al.(1969)]{Bar69} Barnard, A.~J., Cooper, J., \& Shamey, L.~J.\ 1969, \aap, 1, 28 

\bibitem[2004]{CastelliHub2004} Castelli,~F., \&  Hubrig,~D., 2004, A\&A, 425, 263 

\bibitem[2003]{Clarke2003} Clarke, D., 2003, A\&A, 407, 1029

\bibitem[2007]{Decat2007} De Cat, P., 2007, CoAst 150, 167

\bibitem[2008]{Deleuil2008} M. Deleuil, J.C. Meunier, C. Moutou, C. Surace, J.M. Almenara, M. Barbieri, J. Debosscher, H. Deeg, Y. Granet, \&  P. Guterman, 2008, AJ submitted

\bibitem[Dziembowski(1977)]{Dziembowski77}
      Dziembowski, W. 1977, Acta Astron., 27, 95

\bibitem[Dziembowski(1993)]{Dziembowski93}
      Dziembowski, W. A., Moskalik, P., \&  Pamyatnykh, A. A., 1993, MNRAS, 265, 588 

\bibitem[2005]{} Lenz P., \&  Breger M. 2005, CoAst, 146, 53

\bibitem[Gebran et al. (2008)]{2008A&A...479..189G} Gebran, M., Monier, R., \& Richard, O.\ 2008, \aap, 479, 189

\bibitem[\protect\astroncite{{Gonzalez} et~al.}{1995}]{GoLeArMi95}
{Gonzalez}, J.-F., {LeBlanc}, F., {Artru}, M.-C., \&   {Michaud}, G.: 1995,
A\&A, 297, 223 

\bibitem[Grevesse \& Sauval (1998)]{1998SSRv...85..161G} Grevesse, N., \& 
Sauval, A.~J.\ 1998, Space Science Reviews, 85, 161 
	  
\bibitem[Hubeny \& Lanz (1992)]{1992A&A...262..501H} Hubeny, I., \& Lanz, T.\ 1992, \aap, 262, 501 

%\bibitem[]{} Hubrig S., \&  Mathys G., 1995, Com. Ap, 18, 167

\bibitem[]{} Hubrig S., North P., Scholler M., \&  Mathys G., 2006a, AN, 327, 289

\bibitem[]{}{Hubrig}, S., {Gonz{\'a}lez}, J.~F., {Savanov}, I., {Sch{\"o}ller}, M., {Ageorges}, N., {Cowley}, C.~R., \&  {Wolff}, B., 2006, MNRAS, 371, 1953

\bibitem[Iglesias \& Rogers(1996)]{OPAL96}
      Iglesias, C. A., \& Rogers, F. J. 1996, \apj, 464, 943

\bibitem[2004]{koc04} Kochukhov,~O., Bagnulo,~S., Wade,~G.A., Sangalli,~L.,
      Piskunov,~N., Landstreet,~J.D., Petit,~P., \& Sigut,~T.A.A. 2004, A\&A,
      414, 613
\bibitem[]{}{Kochukhov}, O., {Piskunov}, N., {Sachkov}, M., \&  {Kudryavtsev}, D., 2005, A\&A 439, 1093

\bibitem[Kurucz 2005]{2005MSAIS...8...14K} Kurucz, R.~L.\ 2005, Memorie 
della Societa Astronomica Italiana Supplement, 8, 14 

\bibitem[]{} Lanza, A. F., Aigrain, S., Messina, S., et al. 2008, A\&A, submitted

\bibitem[\protect\astroncite{{LeBlanc} et~al.}{2000}]{LeMiRi2000}
{LeBlanc}, F., {Michaud}, G., \& {Richer}, J.: 2000, ApJ, 538, 876

\bibitem[1974]{Preston74} Preston,~G.~W., 1974, ARA\&A, 12, 257 

\bibitem[2001]{RichardMi2001} Richard,~O, Michaud, G., \&  Richer, J. 2001, ApJ, 558, 377

\bibitem[\protect\astroncite{{Richer} et~al.}{1998}]{RiMiRoetal98}
{Richer}, J., {Michaud}, G., {Rogers}, F., {Iglesias}, C., {Turcotte}, S., \&  
  {LeBlanc}, F.: 1998, ApJ, 492, 833

\bibitem[\protect\astroncite{{Richer} et~al.}{2000}]{RiMiTu2000}
{Richer}, J., {Michaud}, G., \&  {Turcotte}, S.: 2000, ApJ, 529, 338

\bibitem[2003]{Robinetal2003} {Robin}, A.~C., {Reyl{\'e}}, C., {Derri{\`e}re}, S., \&  {Picaud}, S. 2003, A\&A 409, 523

\bibitem[2001]{} {Sansonetti}, C. J., \&  {Reader}, J. 2001, Phys. Scr. 63, 219
\bibitem[2002]{shorlin2002} {Shorlin}, S.~L.~S., {Wade}, G.~A., {Donati}, J.-F., {Landstreet}, J.~D., {Petit}, P., {Sigut}, T.~A.~A., \&  {Strasser}, S., 2002, A\&A 392, 637

\bibitem[Smith \& Dworetsky(1993)]{SmithDwo93} Smith, K.~C., \& Dworetsky, M.~M.\ 1993, \aap, 274, 335 


\bibitem[2000]{sti00} Stift,~M.J. 2000, A Peculiar Newsletter, 33

\bibitem[Takeda 1995]{1995PASJ...47..287T} Takeda, Y.\ 1995, \pasj, 47, 
287 

\bibitem[2003]{TurcotteRi2003} Turcotte,~S., \&  Richard,~O, 2003, Ap\&SS, 284, 225

\bibitem[\protect\astroncite{{Turcotte} et~al.}{1998b}]{TuRiMietal98}
{Turcotte}, S., {Richer}, J., {Michaud}, G., {Iglesias}, C.~A., \& {Rogers},
  F.~J.: 1998, ApJ, 504, 539

\bibitem[Vidal et al.(1973)]{1973ApJS...25...37V} Vidal, C.~R., Cooper, J. 
\& Smith, E.~W.\ 1973, \apjs, 25, 37 

\bibitem[2006]{WadeEtal2006} {Wade}, G.~A., {Auri{\`e}re}, M., {Bagnulo}, S., {Donati}, J.-F.,  {Johnson}, N., {Landstreet}, J.~D., {Ligni{\`e}res}, F.,  	{Marsden}, S., {Monin}, D., {Mouillet}, D., {Paletou}, F.,  {Petit}, P., {Toqu{\'e}}, N., {Alecian}, E., \&  {Folsom}, C., 2006, A\&A, 451, 293

\bibitem[W2006]{Wiess2006} Weiss,W.~W. 2006, in ESA Special Publication, vol. 
1306 of ESA Special Publication, 93 (see Baglin, 2006).

\end{thebibliography}
\end{document}